\documentclass[conference,compsoc]{IEEEtran}
\usepackage[]{graphicx}\usepackage[]{xcolor}
\makeatletter
\def\maxwidth{
  \ifdim\Gin@nat@width>\linewidth
    \linewidth
  \else
    \Gin@nat@width
  \fi
}
\makeatother

\definecolor{fgcolor}{rgb}{0.345, 0.345, 0.345}

\usepackage{framed}
\makeatletter
 {\par\unskip\endMakeFramed%
 \at@end@of@kframe}
\makeatother

\definecolor{shadecolor}{rgb}{.97, .97, .97}
\definecolor{messagecolor}{rgb}{0, 0, 0}
\definecolor{warningcolor}{rgb}{1, 0, 1}
\definecolor{errorcolor}{rgb}{1, 0, 0}
\newenvironment{knitrout}{}{}

\usepackage{alltt} 
\usepackage{times}  
\usepackage{helvet}  
\usepackage{courier}  
\usepackage[hyphens]{url}  
\usepackage{graphicx} 
\usepackage{natbib}  
\usepackage{caption} 
\usepackage{tikz}
\usetikzlibrary{positioning}

\usepackage{booktabs}
\usepackage{subfigure}
\usepackage{dcolumn}
  \newcolumntype{d}[1]{D{.}{.}{#1}}
\usepackage{amsmath}

\usepackage{paralist}
\title{The Effects of Group Sanctions on Participation and Toxicity:\\ Quasi-experimental Evidence from the Fediverse}

\author{
    \IEEEauthorblockN{Carl Colglazier\IEEEauthorrefmark{1}, Nathan TeBlunthuis\IEEEauthorrefmark{2}, Aaron Shaw\IEEEauthorrefmark{1}}
    \IEEEauthorblockA{\IEEEauthorrefmark{1}Northwestern University}
    \IEEEauthorblockA{\IEEEauthorrefmark{2}University of Michigan}
}

\IfFileExists{upquote.sty}{\usepackage{upquote}}{}
\begin{document}

\maketitle
\begin{abstract}
 Online communities often overlap and coexist, despite incongruent norms and approaches to content moderation. When communities diverge, decentralized and federated communities may pursue group-level sanctions, including defederation (disconnection) to block communication between members of specific communities. We investigate the effects of defederation in the context of the Fediverse, a set of decentralized, interconnected social networks with independent governance. Mastodon and Pleroma, the most popular software powering the Fediverse, allow administrators on one server to defederate from another. We use a difference-in-differences approach and matched controls to estimate the effects of defederation events on participation and message toxicity among affected members of the blocked and blocking servers. We find that defederation causes a drop in activity for accounts on the blocked servers, but not on the blocking servers. Also, we find no evidence of an effect of defederation on message toxicity.
\end{abstract}

\section{Introduction}

Content moderation in response to toxic and anti-social behavior is pervasive in social media. In general, moderation interventions strive to balance the value of wide and active user bases with the threats posed by conflicts and hate speech \citep{gillespieCustodiansInternetPlatforms2018}.
Websites that host user-generated content and sub-communities apply many kinds of policies and interventions. However, when norms diverge across interconnected, independent communities in the absence of a single (corporate or not) parent or owner, governance and moderation pose acute challenges.

A growing empirical literature has investigated social media content moderation and governance. Moderation actions most frequently target individual posts and accounts, but other group-level sanctions affect entire communities or websites. 
For example, Reddit has banned subreddits and Discord has blocked servers, reducing the prevalence of unwanted behavior within and sometimes beyond the targeted groups \citep{chandrasekharanYouCanStay2017, chandrasekharanQuarantinedExaminingEffects2022, ribeiroPlatformMigrationsCompromise2021, ribeiroDeplatformingDidNot2023, russoSpilloverAntisocialBehavior2023a, zhangGroupSizeIncentives2011}.
Most prior work on group-level sanctions focuses on sanctions applied by central actors such as commercial social media platform staff. However, autonomous community administrators can also enact group-level sanctions such as when a sub-community restricts contributions from members of another sub-community. These decentralized group-level sanctions are distinct in that the  targeted sub-community remains part of the larger network. To our knowledge, the effects of such sanctions remain unexplored empirically.

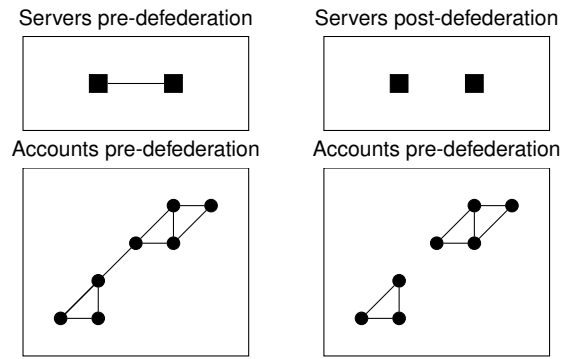
\begin{figure}
\centering
\begin{tikzpicture}[node distance=1.5cm,
    every node/.style={fill=white, font=\sffamily}, align=center]

  \begin{scope}[shift={(0,0)}]
    \draw (0,0) rectangle (3,1.25);
    \node[draw, rectangle, fill=black, scale=1] (S1) at (1,0.625) {};
    \node[draw, rectangle, fill=black, scale=1] (S2) at (2,0.625) {};
    \draw (S1) -- (S2);
    \node[above=5mm, scale=0.8] at (1.5,0.75) {Servers pre-defederation};
  \end{scope}

  \begin{scope}[shift={(4,0)}]
    \draw (0,0) rectangle (3,1.25);
    \node[draw, rectangle, fill=black, scale=1] (S3) at (1,0.625) {};
    \node[draw, rectangle, fill=black, scale=1] (S4) at (2,0.625) {};
    \node[above=5mm, scale=0.8] at (1.5,0.75) {Servers post-defederation};
  \end{scope}

  \begin{scope}[shift={(0,-3)}]
    \draw (0,0) rectangle (3,2.5);
    \node[draw, circle, fill=black, scale=0.5] (A1) at (0.5,0.5) {};
    \node[draw, circle, fill=black, scale=0.5] (A2) at (1,0.5) {};
    \node[draw, circle, fill=black, scale=0.5] (A3) at (1,1) {};
    \node[draw, circle, fill=black, scale=0.5] (A4) at (1.5,1.5) {};
    \node[draw, circle, fill=black, scale=0.5] (A5) at (2,1.5) {};
    \node[draw, circle, fill=black, scale=0.5] (A6) at (2,2) {};
    \node[draw, circle, fill=black, scale=0.5] (A7) at (2.5,2) {};
    \draw (A1) -- (A2) -- (A3) -- (A1) -- (A4) -- (A5) -- (A6) -- (A7) -- (A5);
    \draw (A4) -- (A6);
    \node[above=5mm, scale=0.8] at (1.5,2) {Accounts pre-defederation};
  \end{scope}

  \begin{scope}[shift={(4,-3)}]
    \draw (0,0) rectangle (3,2.5);
    \node[draw, circle, fill=black, scale=0.5] (A1) at (0.5,0.5) {};
    \node[draw, circle, fill=black, scale=0.5] (A2) at (1,0.5) {};
    \node[draw, circle, fill=black, scale=0.5] (A3) at (1,1) {};
    \node[draw, circle, fill=black, scale=0.5] (A4) at (1.5,1.5) {};
    \node[draw, circle, fill=black, scale=0.5] (A5) at (2,1.5) {};
    \node[draw, circle, fill=black, scale=0.5] (A6) at (2,2) {};
    \node[draw, circle, fill=black, scale=0.5] (A7) at (2.5,2) {};
    \draw (A1) -- (A2) -- (A3) -- (A1);
    \draw (A4) -- (A5) -- (A6) -- (A7) -- (A5);
    \draw (A4) -- (A6);
    \node[above=5mm, scale=0.8] at (1.5,2) {Accounts pre-defederation};
  \end{scope}
\end{tikzpicture}
  \caption{Illustration of how defederation disconnects two servers and thereby disconnects the subnetworks of people using each server. The top row shows the network of servers before (left) and after (right) defederation. The bottom row shows the corresponding networks of users. On the left, an edge connects a user on one server with a user on a different server. Defederation (right) disconnects them so they can no longer exchange messages.}
  \label{fig:concept}
\end{figure}

To investigate the effects of decentralized group-level sanctions, we analyze \emph{defederation events} in the Fediverse, a decentralized social media system which consists of independently managed servers that host individual accounts and pass messages using shared protocols. Communication between servers can happen only when the administrators of both servers permit it. Server administrators can revoke such permission by ``defederating'' from (blocking all interactions with) specific servers. Defederation is one of the few tools administrators in a decentralized system have to protect against bad actors or enforce norms from beyond their own servers. 
While many defederation events on the Fediverse occur between servers with no known interaction history, many also come in response to norm violations and toxic interactions across server boundaries with a history of previous interactions. Defederations that cutoff cross-server interactions provide an opportunity to identify the effects of these group sanctions on accounts most likely to be directly affected.

We collect data from 214 defederation events between January 1, 2021 to August 31, 2022 that involved 275 servers and 661 accounts which had previously communicated across subsequently defederated inter-server connections.
Using a combination of non-parametric and parametric methods, we estimate the effects of defederation on two outcomes: posting activity and toxic posting behavior among affected accounts. We find an asymmetric impact on posting activity: Accounts on blocked servers reduce their activity, but not accounts on blocking servers. By contrast, we find that defederation has no effects on post toxicity on either the blocked or blocking servers.

These findings suggest that defederation, although a common group-level sanction on the Fediverse, has mixed effectiveness: Despite the risks of severing communication channels, communities implementing group-level sanctions do not lose activity. This implies that defederation may avoid some of the costs associated with other moderation techniques such as account requirements or group sanctions like geographic blocks \citep{hillHiddenCostsRequiring2021, zhangGroupSizeIncentives2011}.  Although defederation
reduces activity by blocked accounts, we did not find evidence that it made their posts less toxic.  This suggests that defederation may not improve adherence to broadly held norms. 
Our study contributes to knowledge of content moderation on social media in that it (1) describes defederation, a novel form of group-level sanction as instantiated on the Fediverse, (2) derives hypotheses regarding the effects of defederation from prior literature, (3) creates a novel dataset of defederation events, (4) conducts a quasi-experimental analyses to quantify effects of defederation on parties affected on the blocked and blocking servers and (5) finds that defederation has asymmetric effects on activity and no measurable effect on toxicity.

\section{Background}

\subsection{Social media content moderation and governance}

Challenges of decentralized governance and content moderation in networked communication systems are ubiquitous. 
Threats to public safety, trust, and health resulting from toxicity, misinformation, and incivility online are now widely perceived and addressed by various content moderation and governance tools \citep{gruzdTrustSafetySocial2023}. 
National and supra-national governments impose disparate legal regimes such as the GDPR in Europe or the DCMA in the United States.  Simultaneously, corporate platforms employ many strategies, from algorithmic filters to user and content monitoring systems, to monitor and shape content visibility. Leaders of communities such as subreddits and Facebook groups institute their own rules and norms and seek to hold community participants to them \citep{gillespieCustodiansInternetPlatforms2018}.
Interdependence among these components makes effective policymaking, design, and analysis of social media governance especially difficult.

Large corporate platforms and autonomous communities alike must grapple with incongruent norms about online behavior and the appropriate role of the above governance components.   
For instance, today's regulatory assemblage governing social media departs from earlier eras characterized by utopian notions of freedom of expression on the ``electronic frontier''  \citep{kayeSpeechPoliceGlobal2019}. 
Whether governance interventions are understood as ``censorship'' or ``content moderation'' depends not only on one's perspective, but also the on their origin within the institutional hierarchy  \citep{myerswestCensoredSuspendedShadowbanned2018}.
At the highest level, state-led interventions prevent access to sites, domains, services, IP blocks, or protocols \citep{kingHowCensorshipChina2013}. Similarly, network service providers and other intermediaries also engage in blocks or website takedowns justified by terms of service violations or legal risks \citep{vuNoEasyWay2023,ribeiroDeplatformingDidNot2023,hanInfrastructureProvidersThat2022}. Lower in the hierarchy, owners and administrators of specific sites or platforms set their own policies and can enforce them through sanctions applied to individuals as well as whole (sub-)communities \citep{jhaverEvaluatingEffectivenessDeplatforming2021,chandrasekharanQuarantinedExaminingEffects2022, ribeiroPlatformMigrationsCompromise2021}. 
Devolving policy setting and enforcement to increasingly low-level actors poses a trade-off.  Low-level governance can institute local norms and advance a community's specific goals, but may lack the necessary means to prevent important categories of harm. 

A growing literature in the social and computational sciences evaluates the empirical effects of specific content moderation and governance techniques at various institutional levels. Previously studied techniques include deplatforming, quarantines, censorship and geographic blocks, banning, account requirements, and pre-publication filtering \citep{ribeiroDeplatformingDidNot2023, chandrasekharanQuarantinedExaminingEffects2022, kingHowCensorshipChina2013, ribeiroPlatformMigrationsCompromise2021, zhangGroupSizeIncentives2011, hillHiddenCostsRequiring2021, tranRisksBenefitsConsequences2022}. In general, this prior work finds that barriers to specific kinds of contributions from specific kinds of actors cause immediate decreases in the targeted activity among the targeted accounts. As examples, requiring accounts on wikis decreased the quantity of low-quality contributions \citep{hillHiddenCostsRequiring2021}; mainland China's block of Chinese language Wikipedia decreased contributions from editors inside and outside of China \citep{zhangGroupSizeIncentives2011}, and pre-publication moderation increased quality contributions \citep{tranRisksBenefitsConsequences2022}.

Some governance interventions spill-over or have indirect effects (intended or not) beyond their immediate targets. Banning hateful communities on Reddit, for example, reduced speech characteristic of the banned communities by both the active participants in those communities and within the other communities they participated in following the ban \citep{chandrasekharanQuarantinedExaminingEffects2022}. However, when Reddit banned two toxic communities, they subsequently reorganized on other platforms where they experienced declining activity and one had increasing radicalization and toxicity
\citep{ribeiroPlatformMigrationsCompromise2021}.
Banning whole sub-communities may also create spillovers of antisocial and toxic behavior when the former members of banned communities migrate their activities elsewhere \citep{ribeiroPlatformMigrationsCompromise2021, ribeiroDeplatformingDidNot2023}.

The majority of prior work in this area has identified effects of governance and moderation interventions on the largest ``mainstream'' social media platforms such as Facebook, Twitter, or Twitch \citep{jhaverEvaluatingEffectivenessDeplatforming2021, mittsRemovalAntiVaccineContent2022, seeringShapingProAntiSocial2017}. A few have focused on more decentralized or autonomous community environments like subreddits or independently managed wikis \citep{chandrasekharanQuarantinedExaminingEffects2022, hillHiddenCostsRequiring2021}. Many of the interventions studied include group-level sanctions such as political blocks and censorship events affecting whole geographic regions \citep{zhangGroupSizeIncentives2011}, as well as decisions and enforcement actions by ``local'' administrators or moderators \citep{srinivasanContentRemovalModeration2019}. Several more recent studies have considered alternative social media sites, such as alt-right communities Voat, Parler, or Gab, often with an eye towards understanding how interventions on mainstream platforms may have impacted participation in less regulated environments \citep{ribeiroDeplatformingDidNot2023, stockingRoleAlternativeSocial2022}. The current project expands this literature by analyzing the effects of a previously unstudied group-level sanction in a decentralized and federated social media network. 

\subsection{Governance and moderation in federated communication networks}

The architectures of federated communication systems create distinct opportunities and obstacles to governance and moderation. Federated communication networks, such as email, are defined by independently operated servers which let users pass messages both within and across servers. Protocol-based interoperability underpins this design and affords cross-server communication. The resulting server-level autonomy may also offer benefits in terms of scalability and resilience. For instance, new servers may join or leave the network without coordinating with other servers. Such decentralized and complex systems of communication tend to possess a scale-free structure with superior robustness to the failure of single nodes over other classes of large, complex graphs \citep{albertErrorAttackTolerance2000}.

Scalability and resilience are particularly important qualities for communication networks, where the value of the network is often a superlinear function of its size \citep{
vanhoveTestingMetcalfeLaw2016}. 
Expanding a communication network can produce collective benefits to members driven by both the increased space of potential connections (\emph{connectivity}) as well as the increased opportunities for pooled information (\emph{communality}) \citep{fulkConnectiveCommunalPublic1996}. Federated networks also offer profuse opportunities for decentralized action, including modular extensions and localized content and norms \citep{dattaDecentralizedOnlineSocial2010}.

At the same time, the decentralized structure of federated networks also makes it difficult to centrally control or apply uniformity. Technical features typically require a sufficient level of adoption across servers to be useful. 
The overall result is that the user experience in a federated communication network can vary enormously depending on how one connects to it. 

This variability extends to governance. No server can directly dictate the rules or decisions of another. The administrators of each server hold complete autonomous control over their internal affairs. In choosing which server to connect to the network through, individuals also choose a governance regime (intentionally or not). If unsatisfied with one regime, they can (at least in theory) move to another and take their data and social ties with them. Among dissatisfied members of a community, the opportunity for ``exit'' may be especially easy when it costs little time or effort to join another subnetwork \citep{freyEmergenceIntegratedInstitutions2019,freyEffectiveVoiceExit2021}. Individuals wishing to behave anti-socially---to harass, troll, or make offensive posts---may select a server that allows them to do so. Other servers within a federated social media network must address this to protect their own users.

Governance in federated communication networks is not limited to servers' authority over individuals: servers also implement sanctions against each other. If one server hosts anti-social behavior, other servers can implement a server-level intervention. Such interventions, for instance group-level bans or blocks, also happen among semi-autonomous communities within platforms or more centralized environments like Reddit, Wikipedia, and multiplayer video game servers. Yet, to our knowledge, prior work on governance in social media has not analyzed the effects of these sorts of group-level interventions. Doing so offers a valuable opportunity to expand the empirical knowledge in this domain.

\subsection{The effects of community-level defederation as a group sanction}

Servers in a federated network can enact sanctions against other servers in several ways, all of which regulate interactions among their users. For example, servers may filter, limit or slow the propagation of messages from another server through their subnetwork. Of the possible approaches, \textit{defederation}---which completely blocks all communication across server boundaries---is the most extreme. Defederation is widely used when server administrators determine that another server is causing problems to such a degree that managing continued interactions is not worth the trouble. For example, email spam filters often include rules blocking all messages originating from a given email domain or server.

Based on prior content moderation research and the characteristics of defederation as a specific type of governance intervention in decentralized social media, we pursue the following research questions:

\begin{enumerate}
    \item How does defederation impact the activity levels for affected accounts on (a) the \emph{defederated} instance (blocked server); and (b) the \emph{defederating} instance (blocking server)?
    \item How does defederation impact toxic posting behavior among the affected accounts on either (a) the blocked or (b) blocking servers?
\end{enumerate}

When one Fediverse server blocks another, the blocked server no long propagates messages to the blocking server. 
This reduces the size of the audience that can be reached from the blocked server and thereby may decrease the utility of posting on that server.
As with interventions studied in prior research such as bans, quarantines, and deplatforming, we expect that defederation reduces activity among affected accounts on the blocked servers.

The likely consequences of defederation for the blocking server are less clear. Disconnecting from the blocked server decreases the potential audience of the blocking server's users.  As a result,  individuals who lose valuable connections might have fewer reasons to use the network. 
On the other hand, server administrators who initiate defederation likely have an informed rationale. 
They may anticipate the costs and benefits of defederation by observing messages between their server and the servers they consider blocking.
Perhaps more importantly, some volunteer administrators struggle to effectively protect their users from anti-social behavior. 
By enacting defederation, administrators demonstrate competence in server governance, eliciting increased trust and commitment from members of their communities. If users tend to value such competence more than their connections to the blocked server then defederation is unlikely to decrease their activity and could even increase it.

Similarly, it is unclear how defederation will affect toxic posting behavior. Some past studies of deplatforming have found that toxicity increases after some deplatforming events \citep{aliUnderstandingEffectDeplatforming2021,buntainCrossPlatformReactionsPostJanuary2023, chandrasekharanYouCanStay2017}.
Such increases may happen when group-level sanctions provoke blowback or revolts against the imposition of sanctions in the form of non-compliance with sanctions or increasing anti-social behavior \citep{heckathornCollectiveSanctionsCreation1988}. 
However, other studies have found that group-level sanctions decrease or do not change toxicity  \citep{jhaverEvaluatingEffectivenessDeplatforming2021, chandrasekharanQuarantinedExaminingEffects2022}. 

Therefore, we anticipate that defederation will have limited or negligible effects on toxic posting behavior. Accounts on blocked servers may not receive any indication of the reasons for the intervention. While server administrators may choose to publicize their defederation decisions, no mechanism propagates the decisions to the users of targeted servers (much less the users whose behavior may have triggered the block in the first place).  Users on the blocking serer who had interacted across the severed connection likewise experience no other direct effects and retain access to the rest of their communication networks. In the absence of more targeted or focused information about the reasons behind the block, we do not expect it to impact toxic behavior substantially.

\begin{figure}[ht]
\begin{knitrout}
\definecolor{shadecolor}{rgb}{0.969, 0.969, 0.969}\color{fgcolor}
\includegraphics[width=\maxwidth]{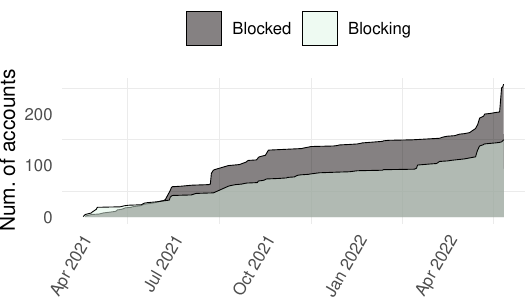} 
\end{knitrout}
  \caption{The y-axis shows the cumulative number of blocked and blocking accounts included in our analysis over our study period.}
  \label{fig:timeline}
\end{figure}

\section{Data, Measures, and Methods}

We pursue an observational, quasi-experimental research design to identify effects of defederation on the activity and toxic posting behavior in the Fediverse. We collected longitudinal trace data from 7,445 publicly listed defederation events and about 104 million public posts that occurred in the Fediverse on either the Mastodon or Pleroma networks between April 2, 2021 and May 31, 2022.
Using this data, we analyze activity of user accounts (for RQ1) and the toxicity of their messages (for RQ2) on the blocking and blocked servers impacted by these events in comparison to matched control accounts. We apply a difference-in-differences approach and present both non-parametric and parametric estimates of the effects of defederation.

\subsection{Empirical Setting}

\begingroup
\renewcommand{\arraystretch}{1.25} 
\begin{table*}[htbp]
    \begin{tabular}{p{0.175\linewidth} | p{0.775\linewidth}}
    \hline
    Term & Description \\
    \hline
    Fediverse & A group of independently operated social media servers which interoperate with shared protocols. \\
    Defederation & A server-to-server block between servers in a federated system. \\
    Mastodon & The most popular and widely used software on the Fediverse. \\
    Pleroma & A lightweight Fediverse software with an API similar to Mastodon and an alternative feature set. \\
    Server administrator & A person or group who controls the technical infrastructure and configuration of a Fediverse server. \\
    Group sanction & A moderation action against a group, not an individual. \\
    \end{tabular}
    \caption{Glossary of key terms.}
\end{table*}
\endgroup

The Fediverse is a decentralized social network comprised of many servers which each institute their own rules, policies, and moderation practices. Each account connects to the network through a home server and can send and receive posts from accounts both within and outside the home server via the ActivityPub protocol. Mastodon and Pleroma are two popular, interoperable ActivityPub-based microblogging systems with overlapping but distinct features. 

Fediverse administrators have nearly complete control over their servers, from choosing and configuring software to setting rules. However, administrators only have direct control over their own server.  Although users of one server can report rule-breaking posts by another server's accounts, only the administrator of an account's home server can  sanction the user, such as by removing the post or sending a warning. Consequently, when two servers come into conflict, administrators must either use persuasion, ignore the problem, or block or filter the other server.

Defederation is the process of severing ties between servers to restrict interactions, typically due to differences in policies, norms, or content. Administrators are responsible for defederation decisions, and they can choose to defederate from specific servers. Defederation helps maintain the autonomy of servers and allows them to adhere to their own values and rules. Individuals with accounts on Fediverse servers do not tend to receive direct notifications about defederation events, though they may either see an announcement by their server administrator. On Mastodon, defederations delete records of the defederated server, meaning that affected accounts lose followers from the defederated server and no longer follow accounts on the defederated server.

\begin{figure}[htbp]
  \centering
  \includegraphics[width=\columnwidth]{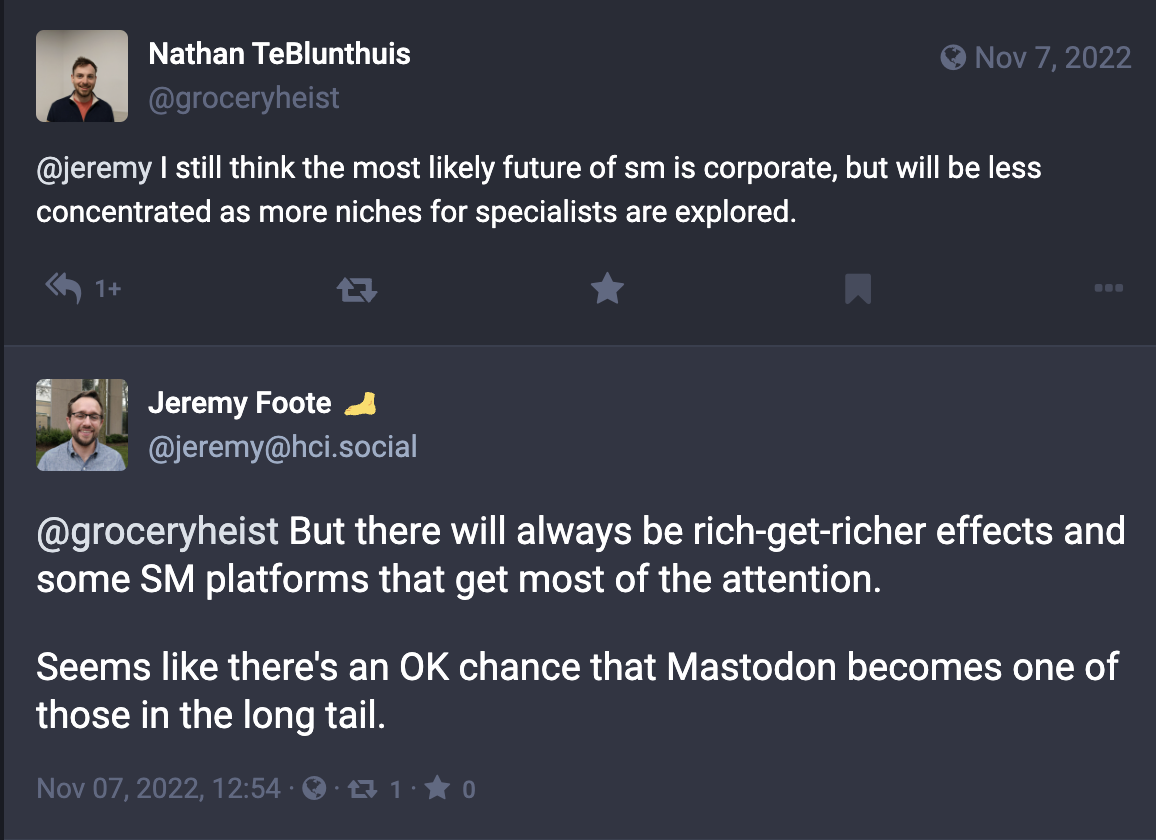}
  \caption{A screenshot shows a cross-server interaction on Mastodon. Note that the account name of the replying user indicates their home server (@\emph{user}@\emph{server}).}
  \label{fig:screenshot}
\end{figure}

\subsection{Data Collection}

We first identify a list of Fediverse servers involved in defederation by compiling a set of known servers which federate with other servers on the list, starting with the set of servers on mastodon.social\footnote{\url{https://mastodon.social/}}. We then iteratively add servers with peer connections to members of this set using an API feature which discloses server-to-server connections. This approach identified 11,025 servers and 7,445 defederation records. We collected data from these servers by running a script daily. The script uses publicly available and documented REST APIs provided by Mastodon and Pleroma to collect timelines of posts. We collected approximately 104 million posts in total.
The post data allows us to identify accounts that engaged in cross-server interaction because usernames indicate the home server.

We identify defederation from the public records servers publish of which other servers they block. We check these records each day and infer the date that defederation occurred by the day a server first appears in the list.  We exclude 472 defederation events where we do not know the exact date of defederation because we are missing a daily snapshot of the record.
We only analyzed defederation events where the public record persisted at least 92 days after the initial event, excluding 1,347 that did not.  Accounts are affected by a defederation event if they had previously sent a reply to an account on the blocked or blocking server. We excluded accounts that may have been affected by defederation events we dropped due to missing data issues.

Our analysis is designed to identify the effects of defederation accounting for trends in activity and toxicity before and after defederation events (defined below).  We therefore analyze data from the 91 days prior to and following each defederation event (dropping the day of the defederation itself). The outcomes of interest are quantities and qualities of posting activity, so we exclude inactive accounts by dropping those that posted fewer than 10 times in the 91 days before a defederation event involving their home server. We also include only accounts that posted at least once prior to the 91 day period and at least once in the final 45 days of this period.

Some servers experience multiple defederation events. In order to avoid potential confounding of our estimates from multiple exposures to defederation, we exclude accounts that experienced defederation events on multiple days.
After dropping some additional accounts in the matching process (described below), the resulting dataset includes 258 accounts whose 71 home servers were defederated and 150 accounts whose 52 home servers defederated another server. This included a total of 214 defederation events.

\subsection{Measures}

We construct measures within the analytic window for all accounts and servers involved in defederations. We also aggregate our measures over 7 day periods (weeks).

\textbf{Defederation events:} We observe defederation events via the records of federation policies published by Mastodon and Pleroma servers within the Fediverse systems we study.
We collected data on 440 blocking servers and 1,136 blocked servers. 
We record the timestamp of each defederation event, which we use to construct a timeline of defederation events.

\textbf{Activity:} The response variable of our analysis for RQ1 is the posting $activity$ of user accounts affected by defederation events.  We measure activity as the number of posts an account makes each week.

\textbf{Toxic behavior:} RQ2 investigates the impact of defederation on toxic posting behavior. We measure post toxicity using the Perspective API model for toxicity, which uses machine learning to predict if a post is a "rude, disrespectful, or unreasonable comment that is likely to make someone leave a discussion" \citep{jigsawPerspectiveAPI2021}. For each post, the model outputs a score ranging from 0--1 corresponding to the estimated probability that the post contains toxic speech. For each account, we aggregated their toxicity during each week by taking the mean toxicity score for their posts during that window.

\textbf{Blocked user accounts:} Defederation is a directional action from one server to another and may not be reciprocated. In our research questions, we estimate the effects of user accounts being blocked from cross-server interactions in a defederation event. We assign accounts to the \emph{blocked treatment group} if \begin{inparaenum}[(1)] \item their home server was defederated (blocked) by another server (the blocking server), \item they replied to an account on the blocking server in the 91 days before defederation, and to ensure the account is active at the time of treatment, \item they post at least once in the last 45 days prior to defederation \end{inparaenum}.  For clarity we refer to the blocked treatment group as $U_0$ below.

\textbf{Blocking user accounts:} In our research questions, we are also interested in effects of defederation on the blocking servers' accounts. We assign accounts on the blocking server to the blocking treatment group ($U_1$) if they meet the parallel conditions to those described above for the blocked treatment group.

\textbf{Time:} As noted above, we aggregate our measures over 7 day periods. We therefore measure ($time$) as the number of such weekly periods before/after the defederation event affecting the account in question excluding the day defederation occurred.

\section{Analytic Plan}

We pursue an identification strategy to estimate the effects of defederation events. For both sets of outcomes---activity levels and toxic posts---we visualize time series, conduct non-parametric statistical tests, and perform regression analysis using  difference-in-differences (DiD) estimators that infer the effects of the treatment on treated accounts from both blocked and blocking servers against a set of matched controls. We use coarsened exact matching to construct this synthetic control group of accounts that are similar to the accounts impacted by defederation (treated accounts).

\subsection{Constructing Matched Control Groups}

All of our analyses depend on comparing user accounts that experienced defederation events with similar accounts that did not. The goal of statistical matching is to reduce bias and account for potential confounders. In this case, we use observable, pre-treatment attributes of accounts that capture how they engage with the wider Fediverse through posting replies to external servers. We match on the following variables: total number of account posts, the number of posts in the 45 days prior to treatment (``post count (45)''), the number of replies sent, the number of other servers engaged with replies, and the number of active accounts on their home server. We applied a log transformation to each of these (highly skewed count) variables. We construct all matching measures over weeks during the 91 days prior to the defederation event experienced by the treated user in question.

We use coarsened exact matching \citep{iacusCausalInferenceBalance2012} with Sturges' rule to determine the number of bins for all variables except the number of active accounts on the server, for which we used four bins to be less strict. We used one-to-one matching, selecting the closest match according to Mahalanobis distance and discarded accounts for which there was not a sufficiently good match (139 for the blocked accounts; 63 for the blocking accounts). Accounts that were treated were not eligible to be matched controls. 

\begin{figure}[htbp]
  \centering

\includegraphics[width=\maxwidth]{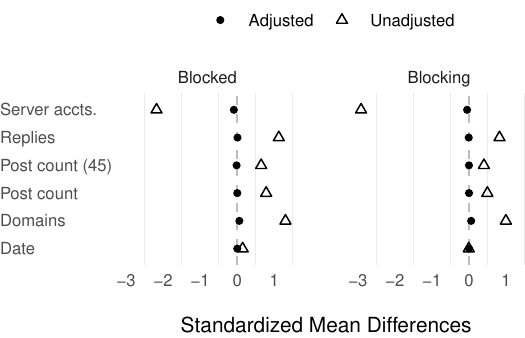} 

  \caption{A covariate balance plot shows the standardized mean difference between treatment and control groups for each measure used in our matching procedure before (unadjusted) and after (adjusted) matching. Our procedure effectively found a group of matched controls similar to the treated accounts along these measures.}
  \label{fig:love}
\end{figure}

We denote the resulting matched control groups as $C_0$, the control group for blocked server users $U_0$, and $C_1$, the control group for blocking server users $U_1$. The blocked accounts had 258 matched units from 397 potential units; the blocking accounts had 150 from 213 potential units. Figure \ref{fig:love} shows the covariate balance using standardized mean differences before and after matching across these groups.

\subsection{Analysis}

Our analysis proceeds in three parts: descriptive comparison and visualization; non-parametric tests; and difference-in-differences estimation. 

\paragraph{Descriptive comparison and visualization:} We calculate descriptive and summary statistics for all measures across the treated and control groups. We also visualize posting activity before and after defederation, plotting the number of weekly posts made by the median user account in each of our four study groups ($U_0$, $U_1$, $C_0$, $C_1$) in the 12 weeks prior to and following defederation with 95\% confidence intervals based on order statistics.  Both the descriptive analysis and the visualization justify and complement our non-parametric and difference-in-differences analyses.

\paragraph{Non-parametric tests:} We first compare changes in activity level and median post toxicity before and after defederation using a non-parametric Wilcoxon signed-rank test, which returns a $W$ test statistic representing the sum of the ranks of the positive differences between paired observation and a p-value which compares to the null hypothesis that the changes between the two groups are zero. This test is robust to outliers, independent of distributional assumptions, and assigns equal weight to each user account irrespective of their activity levels. For toxicity in particular, this non-parametric approach reduces threats of bias due to the opaque machine learning systems that create the values reported from the Perspective API. We conduct this test over all user accounts in both the blocked and blocking groups.

\paragraph{Difference-in-differences (DiD) estimation:} Our primary analysis estimates the causal effect of defederation on the \emph{blocked accounts}  ($U_0$) and the \emph{blocking accounts} ($U_1$) using a difference-in-differences (DiD) framework.  DiD systematically quantifies the changes in a treatment group (i.e., blocked accounts; $U_0$) following an intervention (i.e., a defederation event) relative to changes in a control group (i.e., accounts matched to blocked accounts; $C_0$). 
It does this by modeling temporal trends for each group (treatment; $U_i$ and control; $U_i$) before and after the defederation as well as group-dependent discontinuous ``jumps'' at the moment of intervention. 

Our DiD estimates support causal inferences under several assumptions:
The matching procedure should result in control groups of accounts that were not affected by defederation, but were just as likely to have been affected as the accounts that were, the two groups' trends in the outcome variables should be parallel prior to treatment, and the trends should be linear.
If so, then the difference between the jumps of the two groups quantifies a ``local average treatment effect'' (LATE).\footnote{We say \textit{local} average treatment effect because the estimate applies in specific times and places that may not generalize to other contexts.} We include random intercepts in our model to account for variability between subjects. We fit versions of this model for both outcomes and types of server (blocked and blocking) using the \texttt{brms} R package.

\paragraph{Model for activity:} $Activity$, the number of posts an account makes per week, is an over-dispersed count variable.  We therefore use a negative binomial model formally represented as:

\begin{align*}
y_{i,t}  &\sim NegBinomial(\mu_{i,t}, \phi) \\
\mu_{i,t} &= \beta_0 + \zeta_i + \beta_1 I(i \in U_k) \\
&+ \beta_2 I(t > T) +  \beta_3 t + \beta_4 I(t > T) \cdot t  \\
&+ \beta_5 I(i \in U_k) \cdot I(t > T) + \beta_6 I(i \in U_k) \cdot t \\
&+ \beta_7 I(t > T) \cdot I(i \in U_k) \cdot t
\end{align*}
\noindent Where $y_{i,t}$ is the activity of user account $i$ in week $t$. $T$ indicates the time of defederation, $U_k$ is the treatment group ($U_0$ is the set of blocked accounts and $U_1$ is the set of blocking accounts), $I$ is the indicator function equal to 1 when its parameter is true and 0 if not, and $\zeta_i$ is the random intercept for user account $i$. 
Our parameter of interest is $\beta_5$, the coefficient for post-treatment membership in the treatment group, for testing whether activity increases ($\beta_5 > 0$) or decreases ($\beta_5 < 0$) upon defederation.  The model's other terms are intercepts and trends pre-treatment ($\beta_0$,$\beta_3$) and post-treatment ($\beta_2$,$\beta_4$,$\beta_6$) and membership in the treatment group ($\beta_1$).\footnote{We note that the $\beta_6$ and $\beta_7$ coefficients model the group-specific trends before and after treatment. If these are both 0, this would provide evidence supporting the parallel trends assumption of our DiD models. As shown in Table \ref{tab:did.activity.coefs}, the 95\% credible intervals for both coefficients contain 0, suggesting the assumption is reasonable.}

\paragraph{Model for toxicity:} Our measure for toxicity is calculated at the post level, leading to several key differences in how we model this outcome. Instead of aggregating this at the account level, which would obscure variation, we use a multi-level model with random slopes for group to estimate the effects of defederation as an account-level treatment on post-level toxicity. We use beta regression for our analysis of toxicity because the Perspective API scores are on a range between 0 and 1 and are designed to quantify the probability that a comment is toxic. Our model of toxicity is thus:

\begin{align*}
y_{k,i,t}  &\sim Beta(\mu_{i,t}, \phi) \\
\mu_{i,t} &= K_0 + \zeta_i + \eta_0  (i \in U_k) + \eta_1 (i \notin U_k)  \\
\eta_0 &=\beta_0 + \beta_1 I(t > T) + \beta_2 t  + \beta_3 I(t > T) \cdot t \\
\eta_1 &= \beta_4 + \beta_5 I(t > T) + \beta_6 t + \beta_7 I(t > T) \cdot t
\end{align*}
\noindent where $y_{k,i,t}$ is the toxicity of a given post $k$ by account $i$ in week $t$, $K_0$ is the overall intercept, $\zeta_i$ are the random intercepts for each account, $\eta_0$ are random effects for the treatment group and $\eta_1$ are random effects for the matched control group. 

\section{Results}

\begin{figure}[ht]
\begin{knitrout}
\definecolor{shadecolor}{rgb}{0.969, 0.969, 0.969}\color{fgcolor}
\includegraphics[width=\maxwidth]{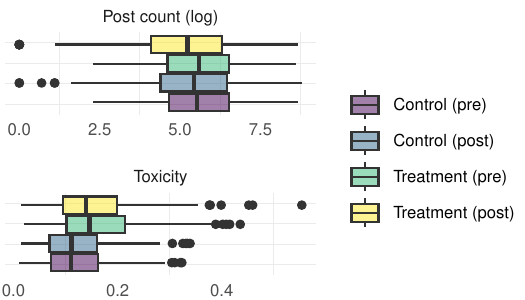} 
\end{knitrout}
  \caption{Box and whisker plots visualize the distributions of our dependent variables within the blocked and blocking groups of user accounts and their matched controls before and after defederation. The lines correspond to the median, the boxes to the inter-quartile range (IQR), the whiskers to the range of the data within 1.5 * IQR, and the dots to data points outside the range of the whiskers.}
  \label{fig:dvplot}
\end{figure}

Figure \ref{fig:dvplot} summarizes our dependent variables across the treatment and matched control groups during the pre- and post-treatment periods. Below, we present results for the two outcomes---activity and toxicity---separately.

\subsection{Effects on Activity}

Research question 1 asks how defederation affected activity among user accounts whose home server either blocked (blocking group) or was blocked by (blocked group) another server. Our visualization of the time series of affected accounts compared to matched controls, non-parametric tests, and difference-in-differences analysis all indicate that blocked group users decreased their activity following defederation while changes among blocking group users were minor or noisy.

\begin{figure}[ht]
\begin{knitrout}
\definecolor{shadecolor}{rgb}{0.969, 0.969, 0.969}\color{fgcolor}
\includegraphics[width=\maxwidth]{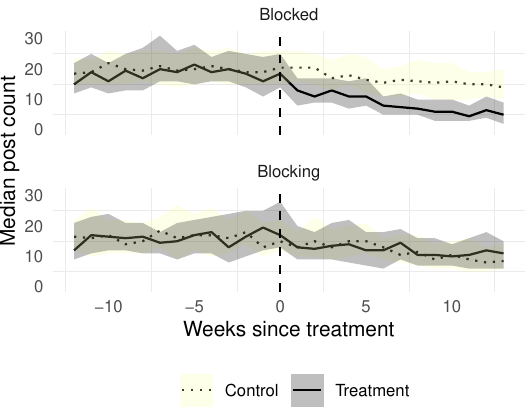} 
\end{knitrout}

  \caption{
    Visualization of activity among blocked and blocking user accounts shows an asymmetric change in activity following defederation. An account with a median post count on the blocked server declines in activity much more rapidly following defederation compared to matched controls while an account with a median post count on the blocking server declines similarly to matched controls.
  }
  \label{fig:median.activity.week}
\end{figure}

Before defederation, the trends and weekly median activity levels remained comparable for accounts on both the blocked and blocking servers. Although all groups reduced activity somewhat after defederation, Figure \ref{fig:median.activity.week} shows the activity levels on the blocked servers quickly diverged from their matched controls after defederation. For instance, while the median post count for treated accounts on blocked servers ($U_0$) was 18.5 posts one week before defederation, this dropped to 13 posts one week after defederation. Compare this to the analogous median post count among matched controls which was 20.5 posts per week both before and after defederation.
In contrast, the median post count among treated accounts on blocking servers ($U_1$) was 17 in the week before and 13 in the week after defederation compared to 15 and 13 for the matched controls ($C_1$).

\subsubsection{Non-parametric tests:}

\begin{table}[htbp]
\centering

\begin{tabular}{lrrr}
\toprule
Group & median & W & p\\
\midrule
$U_0$ & -135.5 & 41197.5 & 0.000\\
$C_0$ & -18.0 & 35762.0 & 0.143\\
$U_1$ & -54.5 & 12413.0 & 0.122\\
$C_1$ & -53.5 & 12520.0 & 0.091\\
\hline
$\Delta_0$ & -39.0 & 39927.0 & 0.000\\
$\Delta_1$ & 3.0 & 10645.5 & 0.421\\
\bottomrule
\end{tabular}

\caption{Non-parametric tests for differences in activity before and after defederation events (summed across all weeks) find a measurable decrease in posting activity for the accounts on blocked servers compared to matched controls but no such change for accounts on blocking servers.} 
\label{table:activity:wilcoxon}
\end{table}

This strong response among the treated accounts on the blocked servers ($U_0$) also appears in the non-parametric test based on post counts summed for all weeks and shown in Table \ref{table:activity:wilcoxon}. Prior to defederation, median activity among blocked accounts and matched controls was similar, yet following treatment, activity for the treated blocked accounts decreased by a median of 135.5 posts (21.9\%) compared to 18 posts (9.5\%) for the matched controls. By contrast, among the user accounts on the blocking servers, we observe very similar post-defederation activity decreases for the treatment and matched control accounts on the blocking server with a decrease in the median of 54.5 posts (10.7\%) by treated accounts compared to 53.5 post (11.1\%) by the the matched controls.
In sum, we observe asymmetric effects of defederation: activity decreases for affected accounts on blocked servers, but not on blocking servers.

\subsubsection{DiD Analysis:}

\begin{figure}[ht]
\centering

\includegraphics[width=\maxwidth]{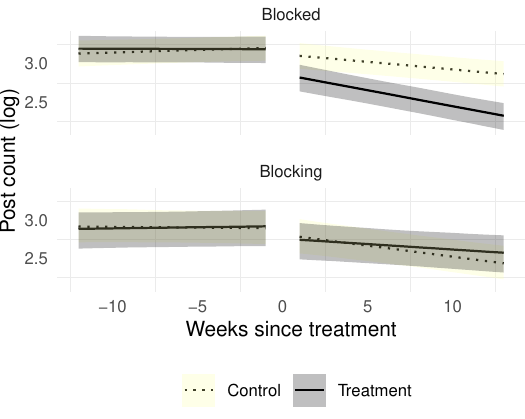} 

\caption{A marginal effects plot visualizes results from our difference-in-differences analysis of account activity by week. We observe a discontinuous activity decrease among accounts on the blocked server that exceeds any decrease in the matched controls, but no corresponding change among accounts on the blocking server. The bands are 95\% credible intervals that account for uncertainty in parameter estimates.}
\label{plot:activity_did}
\end{figure}

Our DiD results corroborate the same pattern: decreased activity among users on the blocked servers, and no change among users on the blocking servers. Table \ref{tab:did.activity.coefs} shows the regression results for both models. Figure \ref{plot:activity_did} plots the modeled activity levels of a median account in each group along with the corresponding 95\% credible intervals quantifying uncertainty in the model's parameter estimates. User accounts on the the blocked servers ($U_0$) decrease their activity compared to matched controls
($\beta_5=-0.24; SE=0.07$). 
For a user account with median pre-treatment activity levels on the blocked server, this corresponds to a reduction from $24.9$ posts per week immediately before treatment to $19.6$ posts per week immediately after treatment (a $21.4$\% reduction), compared to $22.8$ posts per week after treatment for its matched control. 

Consistent with our other results for user accounts on the blocking server,  the 95\% credible interval for the effect of defederation on activity by accounts on the blocking server ($U_1$) contains 0 ($\beta_5=-0.08; SE=0.08$).

\begin{table*}[ht]
\centering

\begin{tabular}{lrrrrrrrr}
\toprule
\multicolumn{1}{c}{ } & \multicolumn{4}{c}{Blocked} & \multicolumn{4}{c}{Blocking} \\
\cmidrule(l{3pt}r{3pt}){2-5} \cmidrule(l{3pt}r{3pt}){6-9}
Term & Estimate & Std error & 2.5\% & 97.5\% & Estimate & Std error & 2.5\% & 97.5\%\\
\midrule
$\beta_0$ (Intercept) & 3.216 & 0.086 & 3.049 & 3.380 & 2.901 & 0.118 & 2.684 & 3.121\\
$\beta_1$ Group & -0.024 & 0.116 & -0.255 & 0.198 & 0.023 & 0.177 & -0.290 & 0.367\\
$\beta_2$ Treatment & -0.089 & 0.045 & -0.178 & 0.002 & -0.080 & 0.060 & -0.201 & 0.034\\
$\beta_3$ Time & 0.006 & 0.004 & -0.002 & 0.015 & -0.002 & 0.006 & -0.014 & 0.009\\
$\beta_4$ Treatment : Time & -0.026 & 0.006 & -0.038 & -0.014 & -0.027 & 0.008 & -0.042 & -0.010\\
$\beta_5$ Group : Treatment & -0.241 & 0.065 & -0.367 & -0.116 & -0.084 & 0.082 & -0.247 & 0.072\\
$\beta_6$ Group : Time & -0.007 & 0.006 & -0.020 & 0.004 & 0.006 & 0.008 & -0.010 & 0.021\\
$\beta_7$ Group : Treatment : Time & -0.015 & 0.009 & -0.031 & 0.002 & 0.009 & 0.011 & -0.013 & 0.032\\
\bottomrule
\end{tabular}

\caption{Difference-in-differences analysis of activity level for user accounts whose server was defederated (blocked group) or whose server defederated another (blocking group).  The 95\% credible interval negative coefficient for membership in the blocked group post-defederation ($\beta_5$) is less than 0, indicating that activity by accounts in this group decreased more than accounts in the matched control group. We do not draw such a conclusion about members of the blocking server because the corresponding credible interval contains 0.}
\label{tab:did.activity.coefs}
\end{table*}

\subsection{Effects on Anti-Social Behavior}

We now present results for our second research question on the effects of defederation on the toxicity of posts by user accounts on the blocked and blocking servers.
Our three analyses: visualization of the time-series of affected accounts compared to matched controls, non-parametric tests, and difference-in-differences analysis are in agreement that defederation did not result in a statistically detectable change in toxicity for affected accounts on either server.

\begin{figure}[ht]
\begin{knitrout}
\definecolor{shadecolor}{rgb}{0.969, 0.969, 0.969}\color{fgcolor}
\includegraphics[width=\maxwidth]{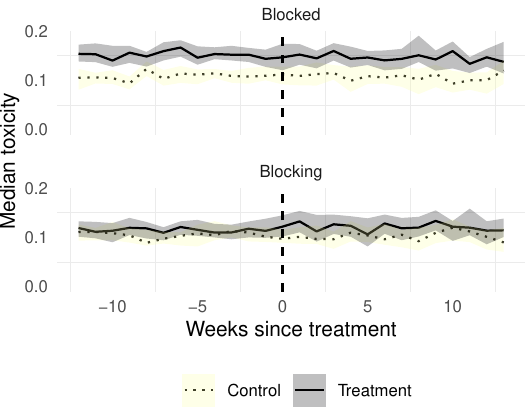} 
\end{knitrout}
  \caption{
    Median toxicity among accounts which posted each week for blocked and blocking user accounts. The median toxicity remained flat for all groups.
  }
  \label{fig:median.toxicity.week}
\end{figure}

The median toxicity plotted in Figure \ref{fig:median.toxicity.week} remained at consistent levels for accounts which posted on a given week for both treatment and matched control groups on both the blocked and blocking servers. The trends remained similar before defederation between the treatment and control groups for accounts on both the blocked and blocking servers.

\subsubsection{Non-parametric tests}

Table \ref{table:toxicity:wilcoxon} summarizes aggregate changes in anti-social behavior (toxicity) after defederation for all groups. Anti-social behavior by user accounts on blocked servers decreases slightly. For example, the median toxicity score for posts from these accounts declined by 0.005, but we lack sufficient statistical power to distinguish this from the null hypothesis that there was no difference in changes between treatment and control groups on the blocked server $(p=0.072)$. We do not observe even such a small change in toxicity among posts from accounts on the blocking server. 

\begin{table}[ht]
  \centering

\begin{tabular}{lrrr}
\toprule
Group & median & W & p\\
\midrule
$U_0$ & -0.006 & 17746 & 0.538\\
$C_0$ & 0.004 & 14000 & 0.950\\
$U_1$ & -0.008 & 6514 & 0.619\\
$C_1$ & 0.001 & 5546 & 0.873\\
\hline
$\Delta_0$ & -0.005 & 17161 & 0.072\\
$\Delta_1$ & 0.000 & 6414 & 0.305\\
\bottomrule
\end{tabular}

  \caption{Non-parametric difference-in-differences for median post toxicity before and after de-federation events. The $W$ test statistic represents the sum of the ranks of the positive differences between paired observations while the p-value compares to the alternative hypothesis that the changes are zero.}
  \label{table:toxicity:wilcoxon}
\end{table}

\begin{table*}[ht]
  \centering

\begin{tabular}{llrrrrrrrr}
\toprule
\multicolumn{2}{c}{ } & \multicolumn{4}{c}{Blocked} & \multicolumn{4}{c}{Blocking} \\
\cmidrule(l{3pt}r{3pt}){3-6} \cmidrule(l{3pt}r{3pt}){7-10}
Group & Term & Estimate & Std error & 2.5\% & 97.5\% & Estimate & Std error & 2.5\% & 97.5\%\\
\midrule
Treatment & $\beta_0$ (Intercept) & -0.143 & 0.438 & -1.197 & 0.763 & -0.042 & 0.284 & -0.754 & 0.518\\
 & $\beta_1$ Treatment & 0.002 & 0.006 & -0.009 & 0.018 & -0.001 & 0.010 & -0.020 & 0.017\\
 & $\beta_2$ Time & 0.001 & 0.001 & 0.000 & 0.002 & 0.001 & 0.001 & -0.001 & 0.003\\
 & $\beta_3$ Treatment : Time & -0.002 & 0.001 & -0.004 & 0.000 & -0.001 & 0.002 & -0.004 & 0.003\\
\addlinespace
Control & $\beta_4$ (Intercept) & 0.136 & 0.436 & -0.907 & 1.044 & 0.048 & 0.283 & -0.613 & 0.649\\
 & $\beta_5$ Treatment & 0.000 & 0.006 & -0.016 & 0.011 & 0.004 & 0.009 & -0.014 & 0.024\\
 & $\beta_6$ Time & 0.001 & 0.001 & 0.000 & 0.002 & 0.003 & 0.001 & 0.001 & 0.005\\
 & $\beta_7$ Treatment : Time & -0.005 & 0.001 & -0.007 & -0.003 & -0.006 & 0.002 & -0.009 & -0.002\\
\bottomrule
\end{tabular}

  \caption{Beta regression coefficients drawn from the posterior of the parametric toxicity DiD model for user accounts whose server was defederated (blocked group) or whose server defederated another (blocking group). For all groups, the 95\% credible intervals for a change in toxicity levels after treatment ($\beta_1$, $\beta_5$) contain 0.}
  \label{tab:did.toxicity.coefs}
\end{table*}

\subsubsection{DiD Analysis}

Again, the results of the DiD analysis for toxicity tell a similar story to the data visualization and non-parametric tests. Our models estimate that toxicity levels of posts from affected accounts on both blocked and blocking servers do not change following defederation events. Full regression results are in Table \ref{tab:did.toxicity.coefs}.
We find no evidence for any significant changes in post toxicity among accounts on either blocked or blocking servers.

\section{Discussion}

We investigated defederation, a group-level sanction where administrators of the blocking server disconnect communication channels between users of their server and the blocked server.
Although one of few actions by which leaders in such networks can protect their community's servers from trolls, harassers, and objectionable content from other servers, defederation's consequences are poorly understood.
Therefore, in a quasi-experimental study, we estimate the effects of defederation on the activity and toxicity of users on both the blocking and blocked servers who had sent or received messages across server boundaries.
Our results provide the first evidence concerning the impacts of defederation in Fediverse communication networks. This evidence can inform governance decisions among Fediverse server administrators and moderators and the design of future decentralized social media networks. 

Research question 1a asked how defederation changed posting activity by affected users on the blocked server. We find that losing communication channels to users on the blocking server causes users on the blocked server who had used such channels to decrease their activity. This helps us understand defederation's potential as an intervention against undesirable behavior. Not only does it protect the blocking server's users from the blocked server, it may decrease the blocked server's value. Defederation may cause a user on the blocked server to lose valued audiences, sources of content, or interpersonal connections and therefore to become less active.

If this positive effect of defederation results from lost connections, defederation might also have a negative consequence: Users of the blocking server may decrease their activity for the same reason. However, our results for research question 1b, which inquired into defederation's consequences on posting activity by affected users on the blocking server, show no measurable effect. These asymmetric effects of defederation on posting activity suggest that administrators may find defederation an effective response to undesirable behavior on other servers without negative impacts on activity by affected user accounts on their own server.

Although sanctions are typically intended to promote normative compliance, group-level sanctions in some prior empirical studies have been met with increases in anti-social behavior or rebellious non-compliance \citep{
mittsRemovalAntiVaccineContent2022, ribeiroPlatformMigrationsCompromise2021, russoSpilloverAntisocialBehavior2023a}.
Therefore, we investigated defederation's effects on toxicity in research question 2. If defederation causes an increase in anti-social behavior, we would expect an increase in toxicity on the blocked server. If defederation causes non-compliance, we would expect toxicity on the blocking server to increase. However, we do not find evidence of either of these adverse indirect outcomes.
In this sense, defederation appears similar to individual and collective sanctions analyzed in prior studies such as account banning, quarantine, and sub-community removal \citep{chandrasekharanQuarantinedExaminingEffects2022, chandrasekharanYouCanStay2017, jhaverEvaluatingEffectivenessDeplatforming2021}.
Some of these group-level sanctions have even improved compliance with widely-held norms \citep{chandrasekharanYouCanStay2017}.  If defederation also does so then we would expect affected users on the blocked server to become less toxic. Yet we observe no such effect.

Among the various kinds of content moderation and governance interventions available to social media systems administrators, server defederation may be among the more extreme group sanctions. However, our analysis suggests that it may also provide administrators with an effective response to undesirable behavior that does not undermine the activity of affected user accounts on the server implementing the block. User accounts affected by defederation events can continue to communicate with others on their respective servers as well as other servers where federated interactions remain permitted.
The disconnection from the network is not total and can respect the local norms that operate among distinct sub-components of the larger network. 
We believe this is a design feature of federated systems that distinguishes them from other kinds of content moderation and governance systems in social media environments. Other, more centralized, platforms may wish to experiment with similar features as doing so may advance user trust and safety without undermining activity.

That said, further research is needed to support recommendations about defederation's overall benefits and limitations in decentralized social media systems.  Such research should uncover the mechanisms that drive the observed changes in activity levels and toxic behaviors, such as changes in user motivations or migration to alternative servers or communication channels. Future work should also investigate how defederation events are perceived and experienced by the people operating affected accounts on both blocked and blocking servers. In addition, insights into the reasons behind administrator decisions to implement defederation blocks as well as the timing of such changes would enrich the findings reported here. As noted above, the defederation events in our study did not occur at random and were likely undertaken with some awareness of the various actors involved.  For these reasons, we encourage caution around the generalizability of our findings beyond the specific context of our study.

Additional limitations of the study relate to the constraints around our data collection, measurement, and analysis strategies. First, our data collection was limited to public status posts, which do not fully capture user interactions within the Fediverse. Private posts and direct messages, which are not accessible through the APIs we used, might exhibit different patterns in response to defederation events. In addition, some servers choose not to publish defederation information. If the effects of defederation emerge anywhere other than public posts, our analysis could not capture them. 

Our analysis also relied on the Perspective API to analyze the content of posts by quantifying their toxicity. While this tool provides a useful proxy, its accuracy may vary across languages and contexts and its errors may affect our statistical results \citep{teblunthuisMisclassificationAutomatedContent2023}. Moreover,  it is conceivable that the contexts for which Perspective designed depart from the Fediverse so substantially that Perspective fails to detect forms of misbehavior affected by defederation. Future research could explore alternative methods for assessing content quality and toxicity. Although language models like those utilized in the Perspective API have significant limitations, we believe that these are likely to be time invariant, at least in the short-run. That is, they would happen on both sides of a defederation event.  In addition, defederation might affect forms of behavior beyond toxicity that future research may investigate such as topics of post or the extent to which they elicit replies or boosts (analogous to retweets). 

In terms of the analysis, our study incorporated several critical assumptions and focused on the effects of defederation events within a relatively narrow timeframe. Matching entails an assumption that selection into treatment or control occurred due to observable variables incorporated into the matching process. A quasi-experimental difference-in-differences analysis similarly assumes parallel trends and comparability across units conditional only on ``as-if'' random exposure to the treatment in question. For example, possible advanced announcements of  defederation events may drive changes in user behavior prior to defederation itself.  While we tried to support such assumptions empirically, they may not hold uniformly, and in addition, the effects of defederation on user behavior might change over time as users adapt to new norms and technologies. Additional and longer duration analyses could address these concerns and shed light on the long-term consequences of defederation events.

\section{Conclusion}

In this study, we investigated the effects of defederation events on the activity levels and toxic posting behavior of accounts in the Fediverse. The results indicate that such events produce asymmetric effects on activity for affected accounts on blocked servers versus those on blocking servers with no increase in toxicity for any groups. The results also highlight the potential of decentralized social networks and their unique mechanisms, such as defederation, in providing communities with tools to manage content moderation and other aspects of online interactions. Future research could explore the causes or reasons behind defederation events, the long-term consequences of defederation, as well as the mechanisms by which defederation produces (asymmetric) effects.  By continuing to study the Fediverse and its affordances, we can better understand how to foster healthy online communities and effective content moderation strategies in a decentralized environment.

\section{Acknowledgements}

The authors would like to thank Ceren Budak and the Community Data Science Collective.
Support for this work was provided by the U.S. National Science Foundation (IIS-191020, IIS-1908850). Additionally, Dr. TeBlunthuis was partially supported by NSF grant IIS-1815875.

\bibliographystyle{aaai24}
\bibliography{Defederation}

\end{document}